\documentclass[12pt,preprint]{aastex}
\usepackage{emulateapj5}
\newcommand{\mincir}{\raise
-2.truept\hbox{\rlap{\hbox{$\sim$}}\raise5.truept\hbox{$<$}\ }}
\newcommand{\magcir}{\raise
-2.truept\hbox{\rlap{\hbox{$\sim$}}\raise5.truept\hbox{$>$}\ }}
\newcommand{\minmag}{\raise
-2.truept\hbox{\rlap{\hbox{$<$}}\raise6.truept\hbox{$<$}\ }}
 
\usepackage{graphicx}

\newenvironment{inlinefigure}{%
\def\@captype{inlinefigure}%
\noindent\begin{minipage}{\linewidth}\begin{center}}
{\end{center}\end{minipage}\smallskip}
\makeatother
\shorttitle{Morphology \& Dynamics of Groups}
\shortauthors{Plionis, Basilakos \& Ragone-Figueroa}

\begin{document}


\title{Morphological \& Dynamical Properties of low redshift 2dFGRS Groups}
\author{M. Plionis\altaffilmark{1,2}, 
S. Basilakos\altaffilmark{1,3}, C. Ragone-Figueroa\altaffilmark{4,5}}
\altaffiltext{1}{Institute of Astronomy \& Astrophysics, 
National Observatory of Athens, Palaia Penteli 152 36, Athens, Greece}
\altaffiltext{2}{Instituto Nacional de Astrof\'{\i}sica \'Optica y
Electr\'onica, AP 51 y 216, 72000, Puebla, Pue, M\'exico}
\altaffiltext{3}{Research Center for Astronomy \& Applied Mathematics,
Academy of Athens, Soranou Efessiou 4, GR-11527 Athens, Greece}
\altaffiltext{4}{Grupo IATE-Observatorio Astron\'omico, Laprida 854, C\'ordoba,
Argentina}
\altaffiltext{5}{Consejo de Investigaciones Cient\'{\i}ficas y
T\'ecnicas de la Rep'/ublica Argentina, C\'ordoba, Argentina}

\begin{abstract} 
We estimate the average group morphological and dynamical characteristics 
of the Percolation-Inferred Galaxy Group (2PIGG) catalogue within
$z\le 0.08$, for which the group space density is roughly constant.
We quantify the different biases that enter in the determination of
these characteristics and we devise statistical correction
procedures to recover their bias free values.
We find that the only acceptable morphological model is that of prolate, or
triaxial with pronounced prolatness, group
shapes having a roughly Gaussian intrinsic axial ratio distribution with
mean $\sim 0.46$ and dispersion of $\sim 0.16$.
After correcting for various biases, the most important of which is a redshift
dependant bias, the median values of the virial mass and virial radius 
of groups with 4 to 30 galaxy members, is: ${\overline M}_v \sim 6 \times 10^{12} \;
h_{72}^{-1} M_{\odot}$, ${\overline R}_v \sim 0.4 \;h^{-1}_{72}$
Mpc, which are significantly smaller than recent literature values that
do not take into account the previously mentioned biases. The group mean crossing time
is $\sim 1.5$ Gyr's, independent of the group galaxy membership.
We also find that there is a correlation of the group size,
velocity dispersion and virial mass with the number of group member
galaxies, a manifestation of the hierarchy of cosmic structures.

\end{abstract}

\keywords{galaxies: clusters: general}

\section{Introduction} 
Groups of galaxies are the lowest level cosmic structures, after
galaxies themselves, in the hierarchy  that leads to the
largest  virialized structures, the clusters
of galaxies. 
There have been a number of recent attempts to construct objectively
selected group and cluster catalogues from magnitude limited 
redshift or photometric galaxy surveys, like the
combination of the Updated Zwicky Catalogue and the Southern Sky
Redshift Survey (UZC-SSRS2), the Sloan digital sky survey
(SDSS), the two-degree field redshift survey (2dFRGS),  the digital Palomar
observatory sky survey; eg.,  Ramella et al (2002); 
Merch\'an \& Zandivarez (2002, 2005); Gal et al. (2003); 
Bachall et al. (2003); Goto et al. (2004); Lee et al. (2004); Lopes et al. (2004);
Eke et al. (2004); Tago et al (2006); Berlind et al (2006). 
Most of the studies that use galaxy redshift information, apply
the so-called friends-of-friends algorithm (FoF) to
the galaxy redshift data by
using a variable linking length that takes into account the drop of
the redshift selection function with increasing redshift. 

It appears that most galaxies are found in groups and
they are therefore extremely important in our attempts to understand
the cosmic structure formation processes.
Since virialization will tend to sphericalize initial anisotropic
distributions of matter, the shape of different cosmic structures is
an indication of their evolutionary stage. Furthermore, the group shape, size
and velocity dispersion are important factors in determining galaxy
member orbits and interaction rates, which are instrumental in understanding
galaxy evolution processes.
Alot of theoretical and observational studies have dealt with the
intrinsic shape of cosmic structures and their dependence on different
cosmological backgrounds, environments, evolutionary stage, etc.  
(e.g., Carter \& Metcalfe 1980; Plionis, Barrow \& Frenk 1991; 
Cooray 2000; Basilakos, Plionis \& Maddox 2000;
Zeldovich, Einasto \& Shandarin 1982;
de Lapparent, Geller \& Huchra 1991;
Plionis, Valdarnini \& Jing 1992; Oleak et al. 1995;
de Theije, Katgert \& van Kampen 1995, Jaaniste et al. 1998; 
Sathyaprakash et al. 1998; 
Valdarnini, Ghizzardi \& Bonometto 1999; Basilakos, Plionis \&
Rowan-Robinson 2001; Jing \& Suto 2002; Kasun \& Evrard 2005; Allgood
et al. 2006; Paz et al. 2006; Sereno et al. 2006).
In the case of groups of galaxies a study of the UZC-SSRS2
group catalogue (Plionis, Basilakos \& Tovmassian 2004)
have shown that they are quite elongated prolate-like systems.

In this paper we use the recently constructed
2PIGG group catalogue (Eke et al. 2004), based on the 2-degree field
redshift survey,
to estimate the group projected and intrinsic shape, size, velocity
dispersion, virial mass and crossing-time 
distributions after correcting for a number of systematic biases.

\section{Group Sample Selection}
The 2PIGG group catalogue (Eke et al. 2004) 
is constructed by applying a friends-of-friends (FoF)
algorithm to the two degree field galaxy redshift
survey (2dFGRS), which contains 191440 galaxies with well
defined magnitude and redshift
selection functions. The FoF
linking parameters were selected after thorough
tests that have been applied on mock $\Lambda$CDM galaxy catalogues.
The resulting 2PIGG group catalogue contains 7020
groups with at least 4 members having a median redshift of 0.11.

The specific group finding algorithm used (Eke et al. 2004)
treats in detail many issues that are related to completeness,
the underlying galaxy selection function and the resulting biases 
that enter in attempts to construct unbiased group or cluster catalogues.

In order to take into account the drop of the underlying galaxy number
density with redshift, due to its magnitude limited nature, Eke et
al. (2004) have used a FoF linking parameter that scales with
redshift. The redshift scaling of this parameter 
is variable in the perpendicular and also 
parallel to the line-of-sight direction, with their ratio being $\sim
11$. 
The necessity to increase the linking volume 
with redshift introduces, however, biases in the
morphological and dynamical characteristics of the resulting groups
which should be taken into
account before attempting to derive their physical and
morphological properties.
For example, an outcome of the above group finding
algorithm is the increase with
redshift of both the velocity dispersion and the projected size
of the candidate groups. Such a systematic effect was also found in
the Ramella et al (2002) group catalogue by Plionis et al (2004). 

In Fig.1 we present the group velocity dispersion and a measure of their
projected size as a function of group redshift (for groups
with membership $n_m\ge 4$). The group projected size, related to the
virial radius, is found by: 
\begin{equation}
r = \frac{n_m (n_m-1)}{2} \left[\sum_{i=1}^{n_m-1} \sum_{j=i+1}^{n_m} 
\frac{1}{D_L \tan (c_{ij} \delta\theta_{ij})}\right]^{-1}\;\;,
\end{equation}
where $D_L$ is the luminosity distance of the group within the concordance
Cosmological model ($\Omega_{\Lambda}=0.7$, $\Omega_{\rm m}=0.3$, 
$h_{72}=0.72$), $\delta\theta_{ij}$ is the angular $(i,j)$-pair 
separation and $c_{ij}$
is an average galaxy pair weight that takes into account the variable 2dFGRS 
incompleteness (in angular, redshift and magnitude limit space). The
individual galaxy weights are taken from Eke et al. (2004).
Note that the above measure of the size is significantly smaller than
the maximum group galaxy-pair separation, $r_{\rm max}$.

A strong redshift dependence is evident in Fig.1. 
The larger velocity dispersion of high-$z$ groups could be possibly
attributed to the fact that we tend to observe at large redshifts
only the richest groups (due to the flux-limited nature of the
2dFGRS). However, the fact that the projected group size and velocity dispersion
increases monotonically with redshift, while the large
and high-$\sigma_v$ groups are not found at lower redshifts, 
can be attributed only to the group identification method (for a
thorough discussion of $z$-dependent effects of the FoF algorithm see
Frederic 1995 and Diaferio et al. 1999). 

\begin{inlinefigure}
\epsscale{1.04}
\plotone{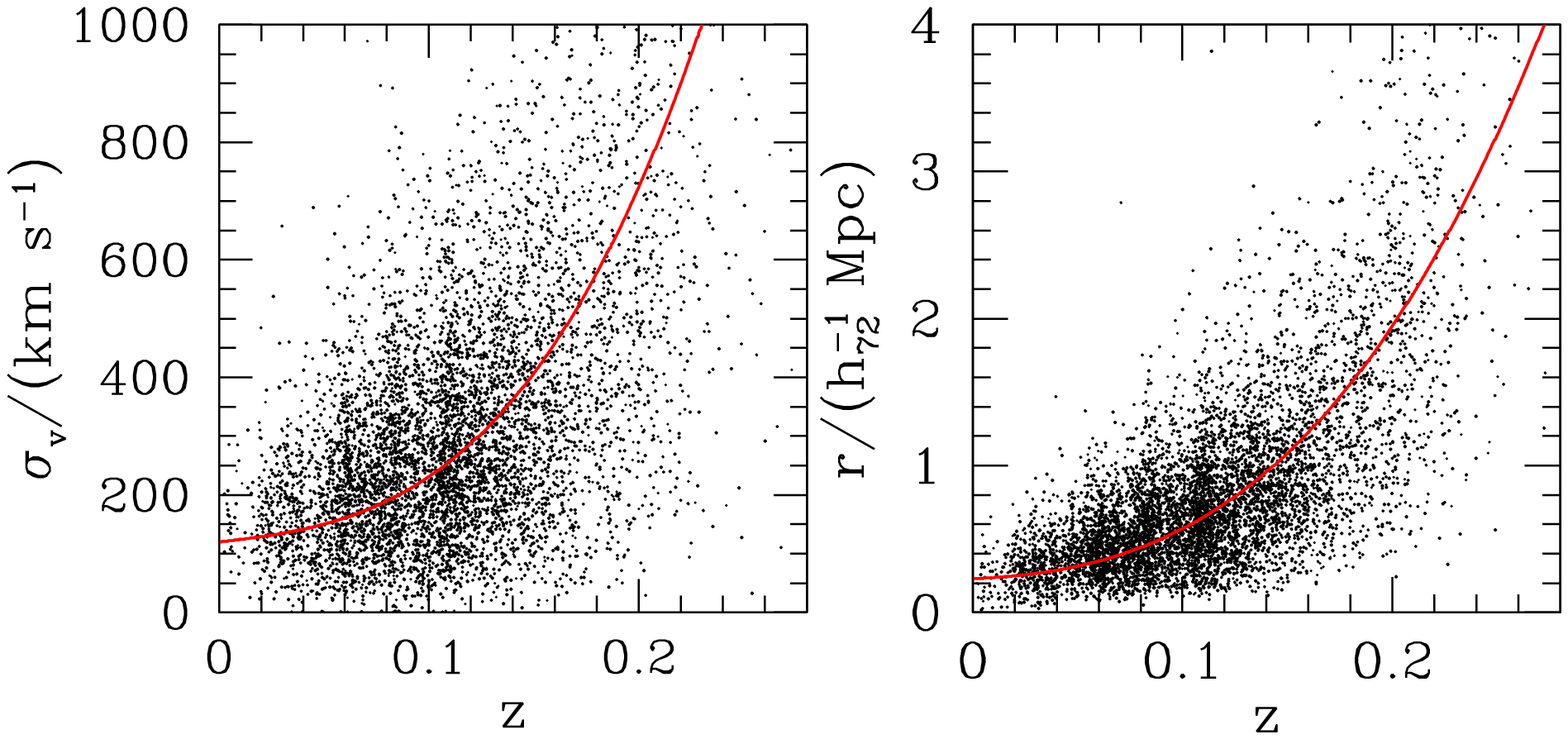}
\figcaption{The dependence of the group velocity dispersion (left panel)
and the group size
(right panel) on redshift. The solid lines are $3^{rd}$ order
polynomial fits.}
\end{inlinefigure}

Therefore, the probability that the groups found are true 
dynamical entities decreases with increasing redshift, 
but even if the high-$z$ groups are real (but contaminated) entities, 
they would constitute a different family of cosmic structures
than the lower-$z$ ones. 
One can attempt to take into account such systematic biases
by applying the same group finding algorithm on N-body 
simulations, the results of which can then be used
to calibrate the statistical results based on the real data
(eg. Eke et al. 2004). 

Since, however, in this study we are interested in deriving the physical
characteristics of the observed groups, we will attempt to quantify
and minimize the various systematic biases. To this end
we also limit the 2PIGG sample to within a redshift 
of $z\simeq 0.08$, a limit within which the density of groups 
(estimated in equal volume shells) is roughly constant, as can be
seen in Figure 2.
\begin{inlinefigure}
\epsscale{0.8}
\plotone{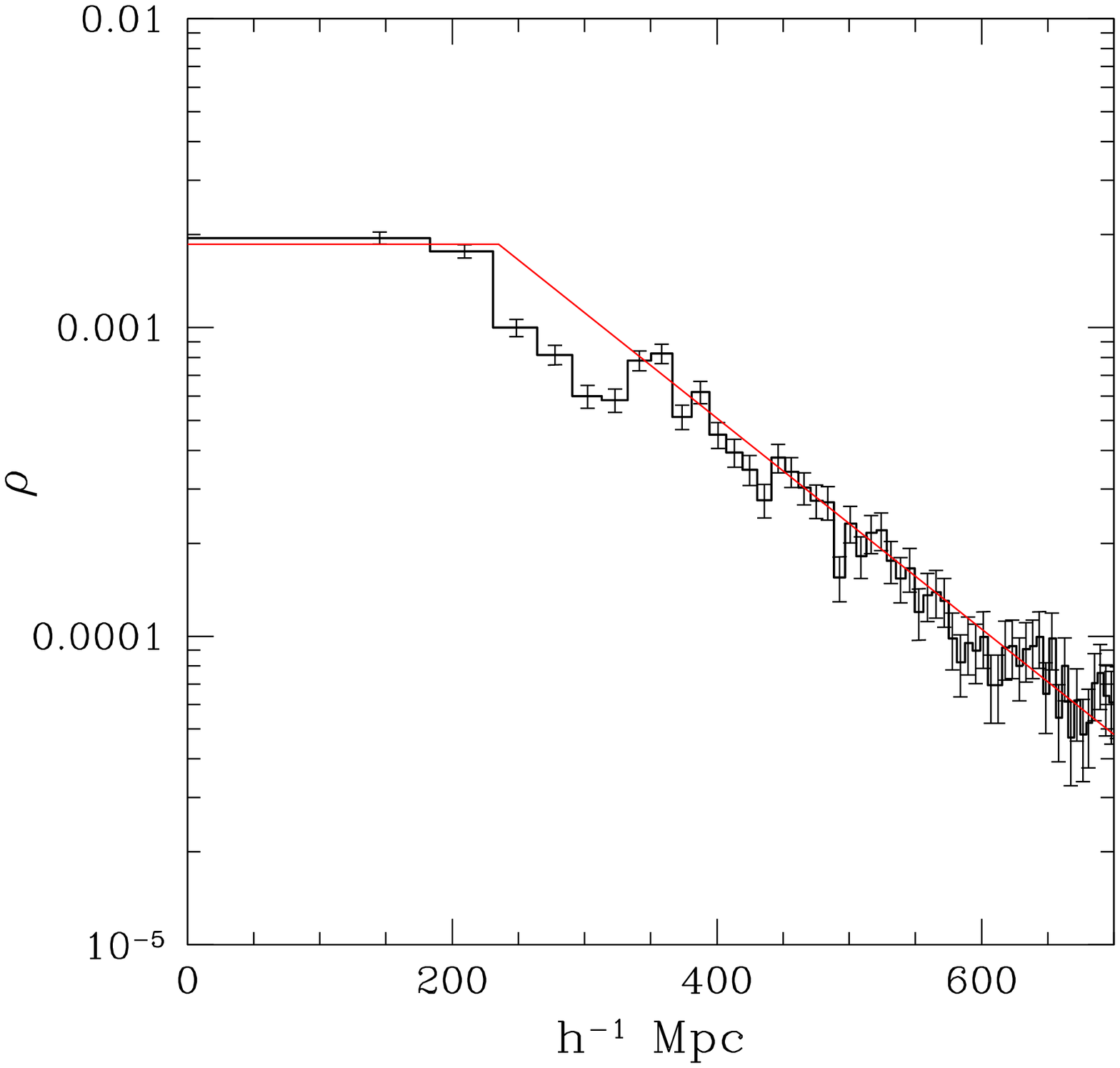}
\figcaption{The group space density 
as a function of distance, estimated in equal volume shells.
The straight line delineates the range where the density is 
roughly constant. The
error-bars are 2$\sigma$ Poisson uncertainties.}
\end{inlinefigure}

We are finally left with 1948 $n_m\ge 4$ 2PIGG groups, out of which 1788
have velocity dispersion estimates from Eke et al. (2004).

\section{Group Projected \& Intrinsic Shapes}
We derive the group projected axial ratio, $q$, by diagonalizing the 
two-dimensional inertia tensor (eg. Carter \& Metcalfe 1983), 
which fits the best ellipse on the projected
discrete distribution of galaxy group members.
In Fig.3 (left panel) we present the group median axial ratio as a function
of group membership (open circles). An interesting trend is apparent
with $\overline{q}$ increasing with increasing $n_m$, while it appears that for
$n_m\magcir 25$ the value of $\overline{q}$ converges to a final value of $\sim 0.56$.
We have verified that this correlation is 
not due to the increase with redshift of the group linking volume, 
which induces the systematic trend seen in Fig.~1.

\subsection{Sampling Effects}
Discreteness effects have been found to affect the group shapes 
by artificially increasing ellipticities with decreasing sampling 
(Paz et al. 2006). We therefore ask the question, what would 
the mean axial-ratio be of groups, having intrinsic flattening that of
the richest 2PIGG groups (ie., $\overline{q}\simeq 0.56$, $n_m\ge 25$),
when sampled with a smaller number of points?
To answer this question we use Monte-Carlo simulations to construct 
a large number ($N_{sim}=30000$ for each $n_m$) of spheroidal 
3D groups having a Gaussian distribution of intrinsic axial ratios,
$\beta$, which we then sample with $n_m$ random points. 
The mean and variance of
the Gaussian is chosen, using a trial and error method, 
such that the corresponding median projected axial
ratio, at the limit of dense sampling, is compatible
with that of the richest 2PIGG groups. This is accomplished 
for a prolate or oblate spheroidal model, with
$\langle \beta\rangle\simeq 0.47$ and $\langle
\beta\rangle\simeq 0.20$, respectively, 
while $\sigma_{\beta}\simeq 0.20$ for both. Note, that since the
Gaussian is bounded between $0\le \beta\le 1$, different values of
$\sigma_{\beta}$ can indeed affect the resulting median projected group
axial ratio.

\begin{inlinefigure}
\epsscale{1.04}
\plotone{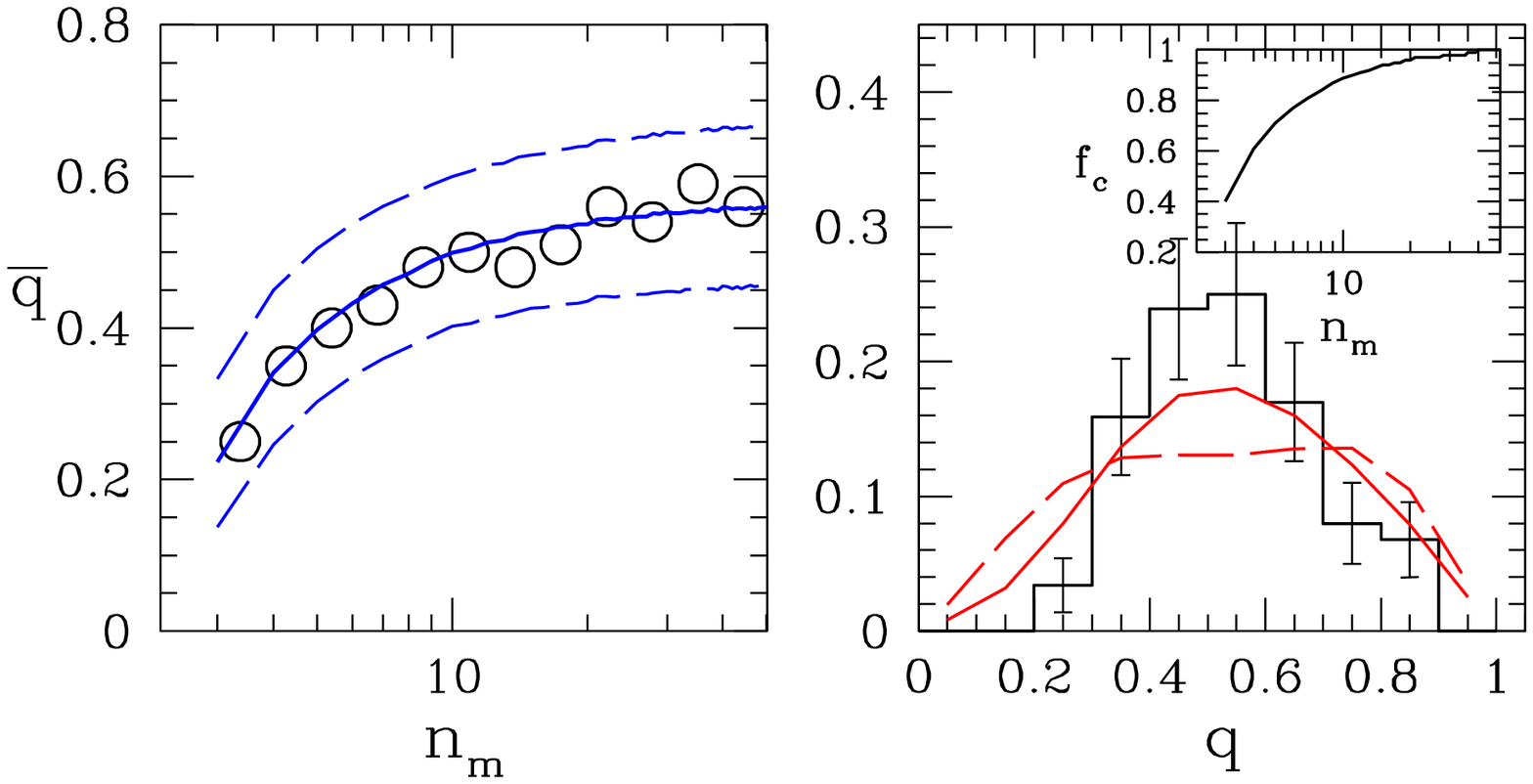}
\figcaption{The correlation between the 2PIGG group ``richness'' ($n_m$) and
 their median projected axial ratio (points). The solid line represent the
expected distribution due to sampling effects (see
text). Broken lines are the 33\% and 67\% quantiles of the corresponding
distribution. The right panel shows the distribution of axial ratios
of 2PIGG groups with $19\le n_m \le 25$ (histogram) and the 
corresponding Monte-Carlo prolate (continuous line) and oblate (dashed
line) groups. The insert shows the correction factor
with which we need to divide the raw group
axial ratios in order to take into account the discreteness bias. 
}
\end{inlinefigure}

Then each group is randomly oriented with respect to the line of
sight, the group members are projected on a surface 
and the projected axial ratio is measured. 
The outcome of the simulations give for $n_m\magcir 20$  
a median value of $\overline{q} \simeq 0.54^{+0.05}_{-0.04}$ (showing 
that a sampling of 20 points per spheroid is adequate to recover the 
input axial ratio). The complete run as a function of $n_m$ can 
be seen as the continuous line in Fig.3 (left panel), from which it 
is evident that the 2PIGG $n_m-\overline{q}$ correlation (open symbols) is 
reproduced exactly (using either the prolate or oblate model for the
Monte-Carlo groups). 
However, the distribution of the projected 2PIGG axial 
ratios was found to be in relatively good agreement only with that
corresponding to the Monte-Carlo prolate groups.
For example, in the
right panel of Fig.3 we plot the axial ratio distribution
of 2PIGG groups with $19\le n_m\le 25$ (histogram) and of the corresponding
Monte-Carlo groups for the two spheroidal models (continuous and
dashed lines).

This proves beyond any doubt that discreetness effects are the cause 
of the observed $n_m-\overline{q}$ correlation and that the 
2PIGG axial ratio distribution is more consistent with that of a prolate
rather that an oblate three-dimensional model.
Therefore, this analysis which serves to clarify the effects of
sampling on the projected group shape, also hints on the intrinsic
morphology of the 2PIGG groups, which formally will be derived in the
next section.

In the insert of the right panel of Fig. 3 we plot the correction factor, $f_c$,
derived from this analysis with which we need to divide the raw group
axial ratios in order to neutrilize the discreteness bias discussed
before. Although we correct accordingly all the group raw axial 
ratios 
we will use the richest ($n_m \ge 10$) 2PIGG groups, for which the discreteness
correction is relatively small, to determine their
intrinsic group shape distribution.
We also put an upper limit to $n_m$
($=30$), in order to exclude from our morphological
analysis clusters of galaxies,
which may have a different dynamical history and thus a different
shape distribution (for cluster shapes see Basilakos, Plionis \&
Maddox 2000 and references therein).

\subsection{Recovering the Intrinsic Axial Ratio Distribution}
An interesting question is whether the group 3D shape distribution can
be inferred from the projected one. This is an inversion
problem for which, under the assumption of random group
orientation with respect to the line of sight and of purely oblate or
prolate spheroidal shapes, there is a unique inversion.
The problem is described by a set of integral equations, 
first investigated by Hubble (1926)
and given by Sandage, Freeman \& Stokes (1970) :
\begin{equation}
f(q)=\frac{1}{q^{2}}\int_{0}^{q}\frac{\beta^{2}\hat{N}_{p}(\beta) 
{\rm d}\beta}
{(1-q^{2})^{1/2}(q^{2}-\beta^{2})^{1/2}} \;\;\; {\rm prolate}
\end{equation}
\begin{equation}
f(q)=q\int_{0}^{q}\frac{\hat{N}_{\circ}(\beta) {\rm d}\beta}
{(1-q^{2})^{1/2}(q^{2}-\beta^{2})^{1/2}} \;\;\; {\rm oblate}
\end{equation}
with
$\hat N_o(\beta)$ and $\hat N_p(\beta)$ the intrinsic oblate
and prolate axial ratio distributions, respectively, and $f(q)$ the
corresponding projected distribution.
The continuous function $f(q)$ is derived from the discrete axial
ratios frequency distribution using the so-called kernel
estimators (for details see Ryden 1996 and references therein).
Although we will not review this method we note that the basic 
kernel estimate of the frequency distribution is defined as:
\begin{equation}
\hat{f}(q)=\frac{1}{Nh} \sum_{i}^{N}\ K\left(\frac{q-q_{i}}{h}\right) \; \; ,
\end{equation}
where $q_{i}$ are the group axial ratios and
$K(t)$ is the kernel function (assumed here to be a Gaussian), 
defined so that $\int K(t) {\rm d}t=1$,
and $h$ is the ``kernel width" which determines the balance between
smoothing and noise in the estimated distribution.
The value of $h$ is chosen so that the expected value of the
integrated mean square error between the true, $f(q)$, and estimated, 
$\hat f(q)$, distributions 
is minimised (eg., Vio et al. 1994; Tremblay \& Merritt 1995).
Inverting then the above equations
gives us the distribution of true axial ratios as a function of $\hat{f}(q)$
(eg. Fall \& Frenk 1983):
\begin{equation}\label{eq:oblate}
\hat{N}_{o}(\beta)=\frac{2\beta (1-\beta^{2})^{1/2}}{\pi} \int_{0}^{\beta}
\ \frac{\rm d}{{\rm d}q}\left(\frac{\hat{f}}{q} \right)\frac{{\rm d}q}
{(\beta^{2}-q^{2})^{1/2}}
\end{equation}
and
\begin{equation}\label{eq:prolate}
\hat{N}_{p}(\beta)
=\frac{2(1-\beta^{2})^{1/2}}{\pi\beta} \int_{0}^{\beta}
\ \frac{\rm d}{{\rm d}q}(q^{2}\hat{f})
\frac{{\rm d}q}{(\beta^{2}-q^{2})^{1/2}} \; \; .
\end{equation}
with $\hat{f}(0)=0$. 
Integrating numerically eq.(\ref{eq:oblate})
and eq.(\ref{eq:prolate}) allowing $\hat{N}_{p}(\beta)$ and
$\hat{N}_{o}(\beta)$ to take any value, we can derive the inverted
(3D) axial ratio distributions. If groups are a mixture of
the two spheroidal populations or they are triaxial ellipsoids
there is no unique inversion (Plionis, Barrow \& Frenk 1991). However, all may
not be lost and although the exact shape distribution may not be
recovered accurately one could possibly infer whether the 3D halo
shapes are predominantly more prolate or oblate-like. 

\subsection{Testing the method with N-body simulations}
Here we attempt to investigate the accuracy of the shape inversion
method when the intrinsic 3D shape distribution is not that of pure spheroids.
To this end we use GADGET2 (Springel et al. 2005) to 
run a large ($L=500 \; h^{-1}$ Mpc,
$N_{p}=512^3$ particles) N-body (DM only) simulation of a 
flat low-density cold dark matter model with
a matter density $\Omega_{\rm m}=1-\Omega_{\Lambda}=0.3$, Hubble 
constant $H_{\circ}=72$ km s$^{-1}$ Mpc$^{-1}$ and a normalization parameter
of $\sigma_{8}=0.9$ 
The particle mass is $m_{\rm p}\ge 7.7\times
10^{10}\,h^{-1}\,M_{\odot}$ comparable to the mass of one single
galaxy. 
The halos are defined using a FoF algorithm with a linking
length given by $l=0.17\langle n\rangle^{-1/3}$ where $\langle n \rangle$ 
is the mean density. This linking length corresponds to an overdensity 
$\simeq 330$ at the present epoch ($z=0$). We will
use intermediate richness halos with
$6\times 10^{13}h^{-1}M_{\odot} \le M_{\rm h} \le 8\times
10^{13}h^{-1}M_{\odot}$, for which there are more than 700 particles
per halo and which therefore are free of discreteness effects. 
The total number of such halos is 2610.

Dark matter halo shapes are 
quantified using the
so called triaxiality index (Franx, Illingworth, G., de Zeeuw 1991), 
defined as: $T=(a^2-b^2)/(a^2-c^2)$ (with $a, b, c$ are the major,
intermediate and minor halo axes of the best fit ellipsoid) which has 
limiting values of $T=1$ (prolate spheroid) and $T=0$ (oblate spheroid).
Our results, which are in agreement with other
studies,  show that the fraction of halos with
pronounced prolatness (ie., large $T$s) is significantly higher than that of
oblate-like halos. Overall we obtain from our simulated halos that 
$\langle T \rangle\simeq 0.73$.

\begin{inlinefigure}
\epsscale{1.04}
\plotone{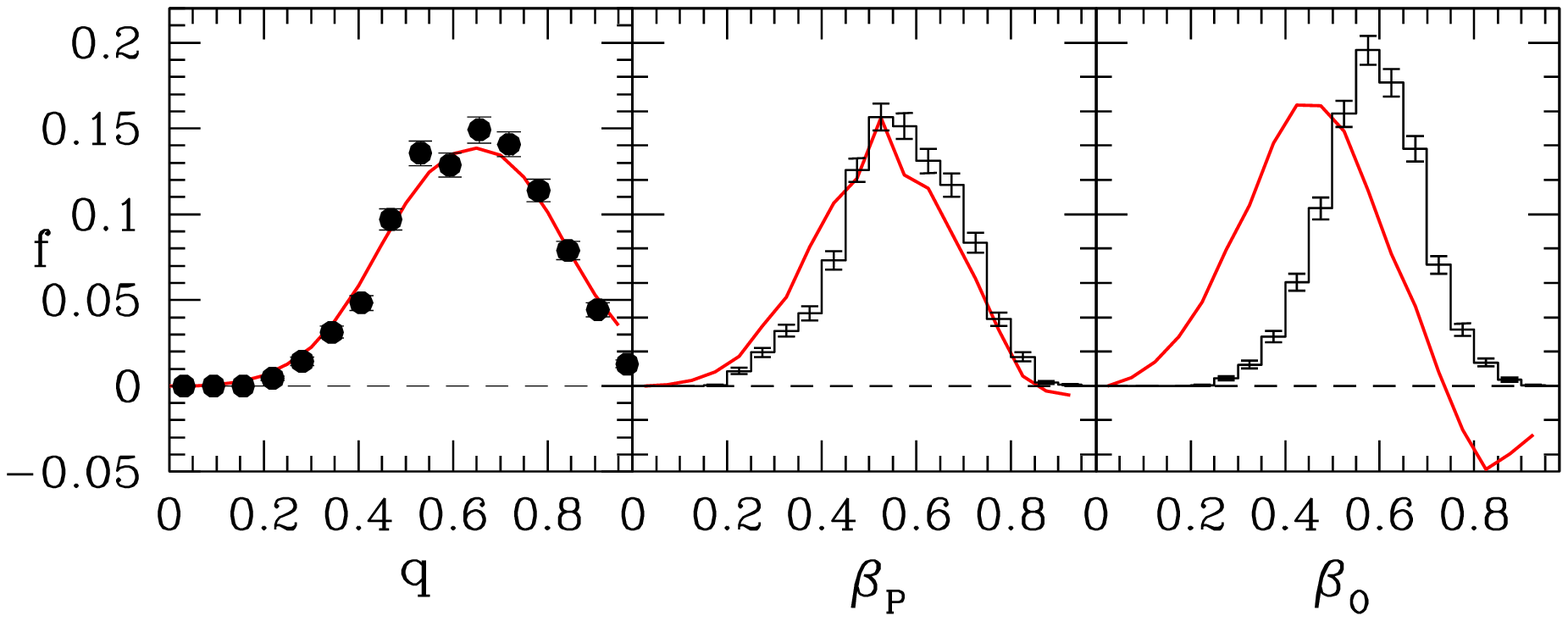}
\figcaption{
{\em Left panel:} The projected halo axial ratio distribution 
(points) and the smooth fit from the nonparametric 
kernel estimator (solid line).
{\em Central \& right Panels:} Comparison of the inverted intrinsic halo axial
ratio distribution (continuous line) with the
distribution of ``average'' spheroidal fits to the 3D halos (histograms), 
either for the prolate (central panel) or oblate (right panel) models.}
\end{inlinefigure}

We now project into 2D the distribution of halo particles and
determine their projected axial ratios 
following the same procedure as in the real group data. 
In Fig.4 (left panel) we present the discrete and continuous - $f(q)$ -
distributions of the projected in 2D halo axial ratios.
In the central and right panels 
we present the inverted 3D axial ratio distributions 
(continuous lines) for the prolate and oblate models, respectively. 
It is evident that the inverted oblate-model distribution is
unacceptable due to the many negative values (at large $\beta$'s), 
while the opposite is true for the prolate-model distribution. 

As a further test,  we plot as histograms
the intrinsic axial ratio distribution of the ``average'' prolate or
oblate spheroidal fits to the 3D halos. 
These fits are realized by estimating the corresponding axial ratios
by $\beta_{P}=(b+c)/2a$ and $\beta_{O}=2c/(a+b)$.
It is evident that the purely
oblate model fails miserably to even come close to the inverted
distribution while the prolate model fits relatively well the
corresponding inverted 3D prolate-model distribution. 
These results are in accordance with the intrinsic
halo shapes determined in 3D, which were found to have a higher
prolatness, as discussed previously.

We therefore conclude that applying the previously discussed
inversion method to observational data we can infer, even in the 
event of triaxial
ellipsoidal halo shapes, the dominance of prolate or oblate-like 3D
shapes, if such does exist.

\subsection{2PIGG Intrinsic Shape}
In Figure 5 (left panel)
we present the raw and discreteness corrected
projected axial ratio distributions for the 2PIGG
groups with $10\le n_m\le 30$ (circles) with their Poisson 1$\sigma$ 
error bars, while the 
solid lines shows the continuous fits, $f(q)$.
The median discreteness corrected axial ratio is $\overline{q}
=0.55\pm 0.08$, while the uncorrected one is $\sim$0.51.

\begin{inlinefigure}
\epsscale{1.04}
\plotone{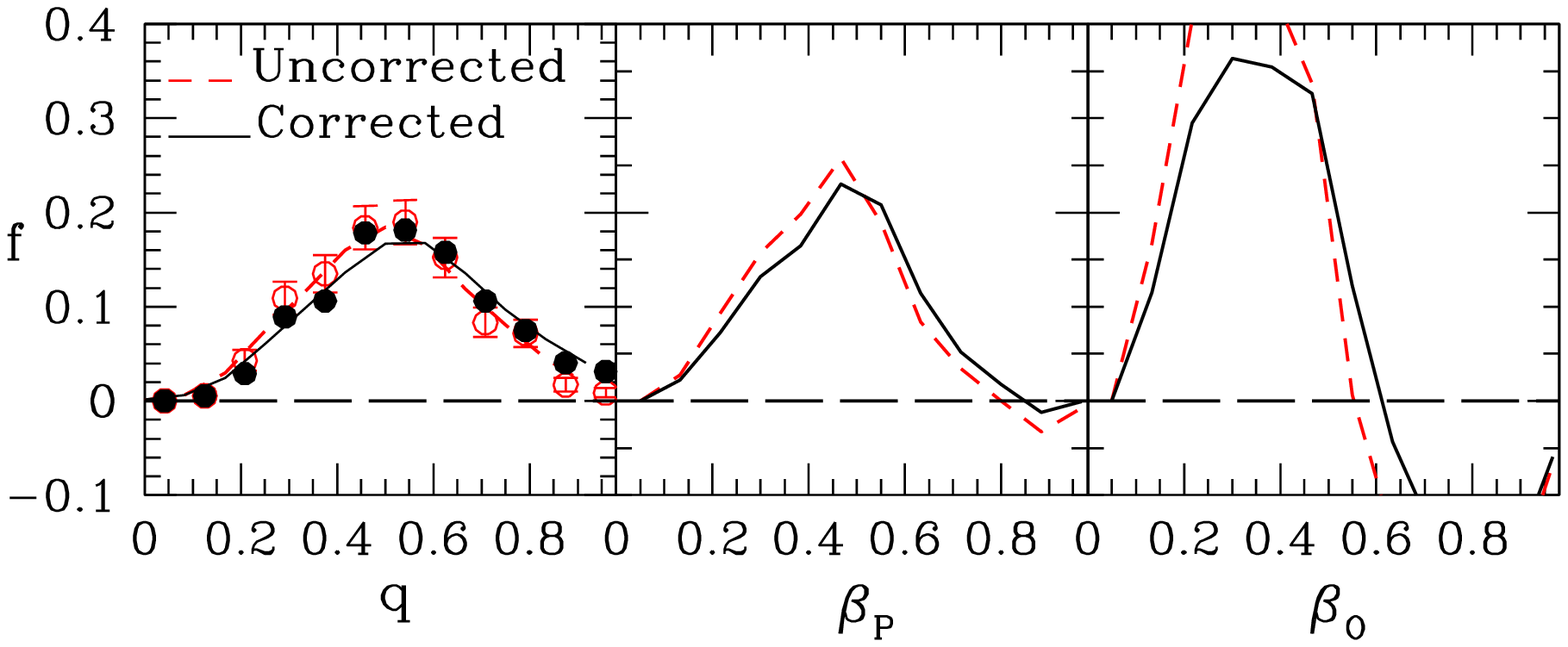}
\figcaption{
{\em Left panel:} The 2PIGG projected group axial ratio distribution 
(points) and the smooth fit from the nonparametric 
kernel estimator (lines), for the discreteness corrected case
(filled points and continuous line) and the uncorrected case (open
points and dashed line).
{\em Central \& right Panels:} The inverted intrinsic halo axial
ratio distribution for the prolate (central panel) and oblate (right
panel) cases.}
\end{inlinefigure}

In the central and left panels of Figure 5
we present the inverted intrinsic axial ratio distribution.
The oblate model is completely
unacceptable since it produces many negative values (at large
$\beta$'s), while the prolate model fairs quite-well, providing a
roughly Gaussian intrinsic axial ratio distribution with $\langle \beta \rangle
\simeq 0.46$ and $\sigma_\beta\simeq 0.16$ (which are also in very good
agreement with the results of the Monte-Carlo procedure of section 3.1).
 Taking into account also
the N-body simulation analysis of section 3.3, our results
imply that the 2PIGG group shapes is well
represented only by that of triaxial ellipsoids with a pronounced prolatness,
which is also in agreement with the previous
analysis of poor groups (Plionis et al. 2004),
and compact groups (eg. Oleak et al. 1995). 

\section{Group dynamical characteristics}
We are now interested in determining the typical size, velocity
dispersion, crossing time and virial mass of the 2PIGG groups. 
We remind the reader of the strong $z$-dependence of these parameters
(Fig.1 ) which is due to the group finding algorithm. Although, we
have limited our analysis to 2PIGG groups within $z=0.08$,  
we observe that even within this $z$-limit, the previously
discussed redshift dependant bias, although relatively weak,
is still present (between $z\simeq0$ and $z=0.08$ the 
average values of $\sigma_v$ and $r$ 
increase by $\sim 50\%$).
In order to statistically correct for this bias
we fitted third order polynomial functions, $V(\sigma_v,z)$ and
$S(r,z)$, to the data (solid curves in Fig.1) and then corrected 
the raw $\sigma_v$ and $r$ values by weighting them with
$V(\sigma_v,0)/V(\sigma_v,z)$ and $S(r,0)/S(r,z)$, respectively. 
However, we present results also for the very local volume ($z\le
0.03$) which apparently is unaffected by the redshift dependent bias
(see Table 1).
\begin{inlinefigure}
\epsscale{1.04}
\plotone{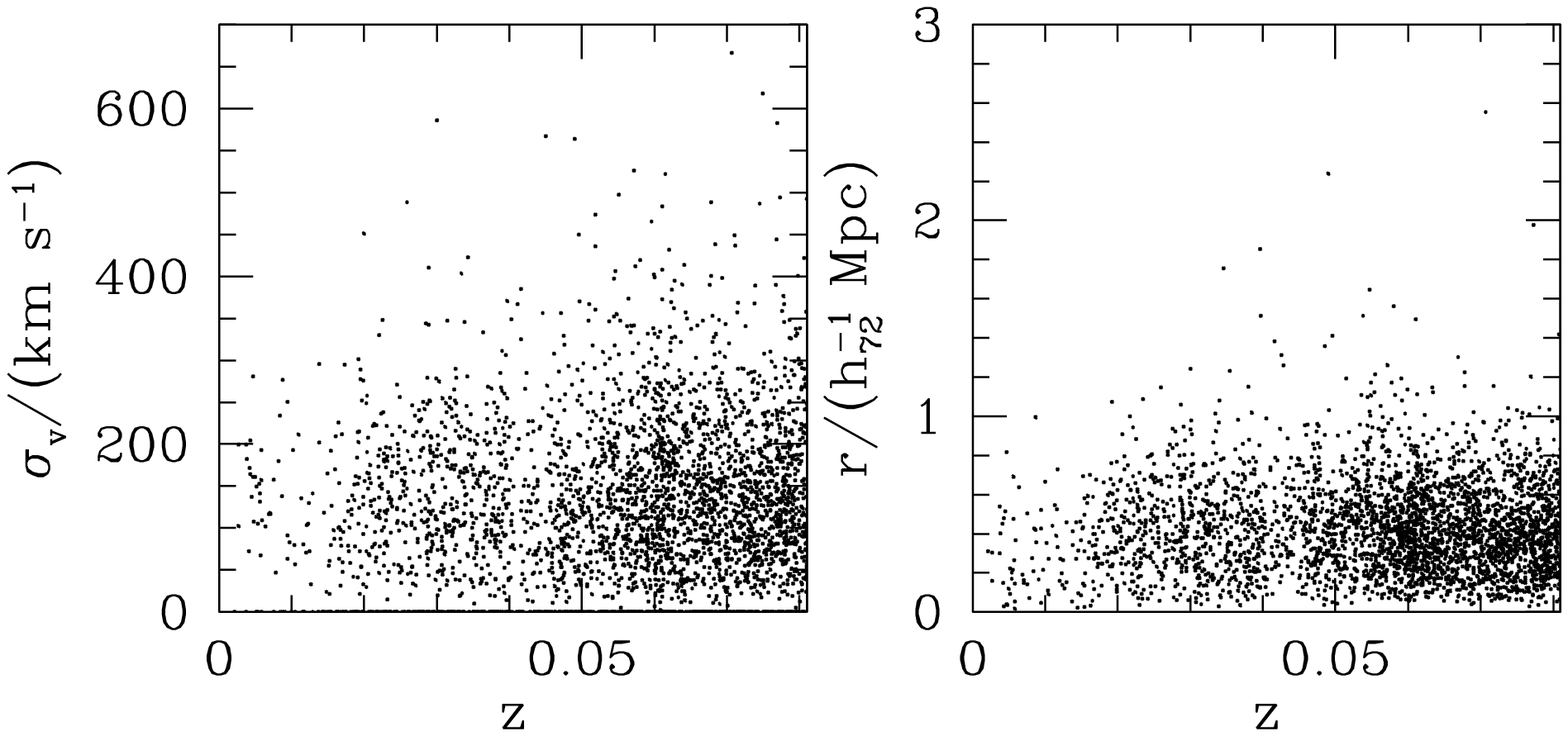}
\figcaption{
The dependence of the corrected group velocity dispersion (left panel)
and group size (right panel) on redshift.}
\end{inlinefigure}

Also, due to the discreteness effects, quantified in section 3.1,
there is an enhancement of the projected group size, $r$, especially at small
$n_m$'s (eg. we find a $\sim 18\%$ increase of the projected
size for groups with $n_m=4$).
For the velocity dispersion case we sample a Gaussian
velocity distribution with $n_m$ and we find a generally small effect
(eg. there is a $\sim 10\%$ reduction of $\sigma_v$ for groups with
$n_m=4$). In Fig.6 we present the corrected, for all the above biases, 
group velocity dispersion and size as a function of redshift for our
$z\le 0.08$ sample. It is evident that the strong dependence on
redshift has been eradicated and although the individual values of
velocity dispersion and size may diverge from the intrinsic values,
the overall population statistics should be correct.

After applying the above corrections to the group velocity dispersion
and size, we can estimate their virial mass and crossing time by: 
\begin{equation}
M_v=\frac{3 \sigma_v^2 R_v}{G}\;\;\;,\;\;\;\;\;
\tau=\frac{R_v}{\sqrt{3} \sigma_v}\;,
\end{equation}
where the group virial radius is $R_v=(\pi/2)\;r$ (with
$r$ given by eq.1). In Table 1 we present the median values of these
parameters, as well as the maximum projected group inter-galaxy separation
($r_{\rm max}$), for the bias corrected sample (with $z\le 0.08$), for the
unaffected by the bias very local sample ($z\le 0.03$), as well as 
for two different redshift-limited samples, but with no corrections 
applied in order to appreciate the extend of the biases
(note that the dominant correction is by far that of the redshift
dependence). The comparison of these results yields that:

\noindent
(a) the effect of correcting the $z\le 0.08$ groups is appreciable,
while the corrected median values of $M_v$ and $R_v$ are about 
equal to those of the local ($z\le 0.03)$, unaffected by biases,
sample, and

\noindent
(b) the effect is extremely large when using groups of any redshift (the
uncorrected median values of $M_v$ and $R_v$ are respectively a 
factor of $\sim$8 and $\sim$3 larger than the corresponding corrected
ones).

When we now compare with determinations from other group catalogues, 
identified using similar FoF based algorithms, we can appreciate the extent of
the effect (see Table 1 of Merch\'an \& Zandivarez 2004). The typical
median values, resulting from different group catalogues that do not
take into account this effect, are: ${\overline M}_v \simeq 5.5 \times
10^{13} \; h^{-1}_{72} \; M_{\odot}$ and ${\overline R}_v \simeq 1.4 \; 
h^{-1}_{72}$ Mpc. In order to avoid such effects Tago et
al. (2006) choose to use a constant in redshift FoF linking
parameter. However, the
decrease of the redshift selection function of the 2dF parent galaxy
catalogue implies that they select intrinsically different 
type of groups as a function of redshift (ie., at higher
redshifts they will tend to select more centrally condensed 
groups or the centers of clusters that have a high enough central
density to survive the drop of the selection function at their distance).
\begin{inlinefigure}
\epsscale{1.0}
\plotone{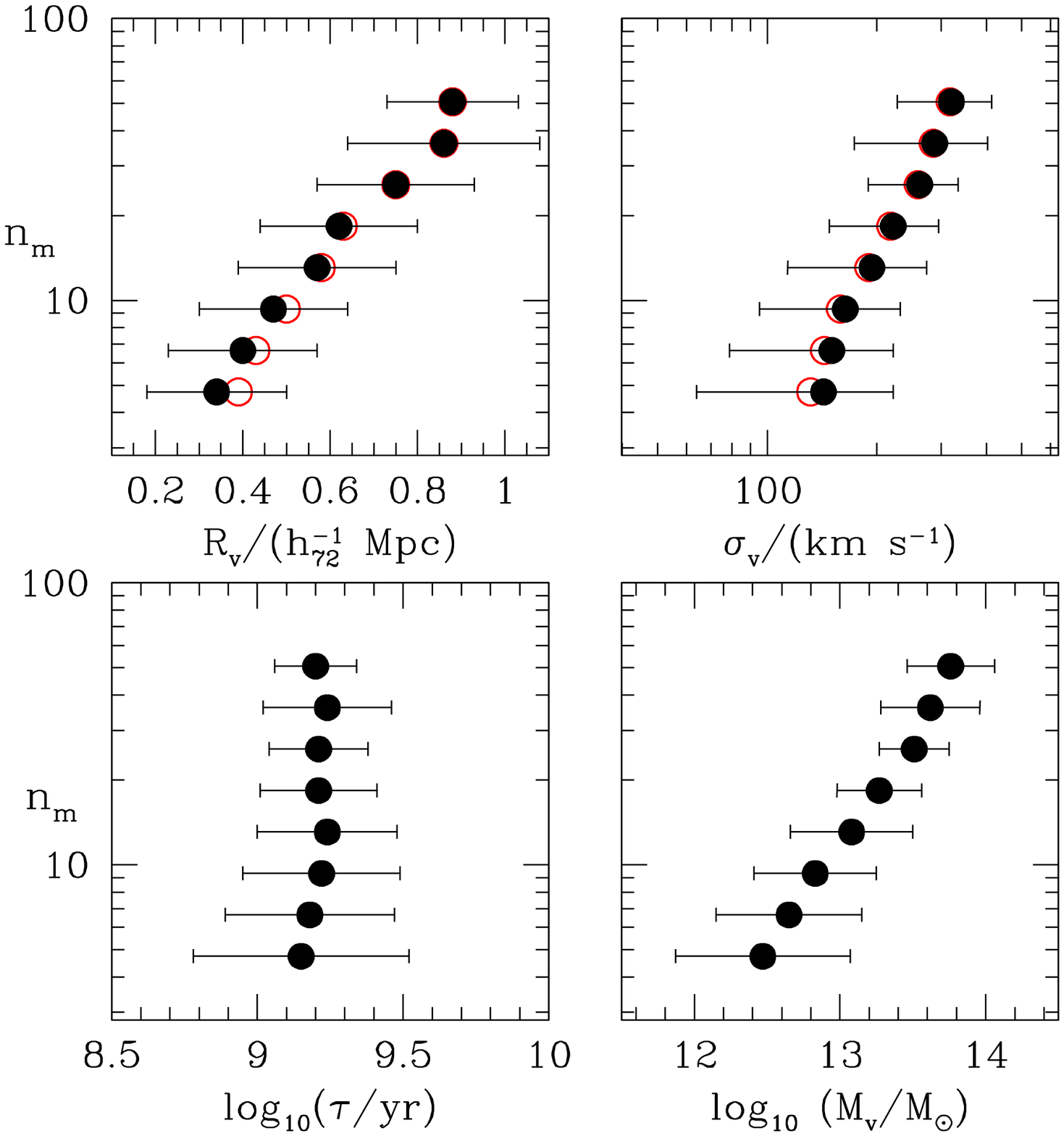}
\figcaption{2PIGG group virial radius (upper left panel), 
the velocity dispersion (upper right panel), the corresponding
group crossing time (lower left) 
and the group virial mass (lower right) as a function of
group ``richness'' ($n_m$). 
The open circles correspond to the raw values and filled circles to
the discreteness corrected values. Note, that here we allow also
$n_m>30$ groups.}
\end{inlinefigure}

It is important to note that although the presented median values of the
various dynamical group parameters are useful in order to compare with other
studies, they are rather ill-defined, since 
we are mixing groups of different richness which could
have distinct morphological and dynamical characteristics. Indeed,
this is the case and 
in Fig.7 we present the correlation between various dynamical and
morphological group parameters 
with group ``richness'' ($n_m$). It is
evident that 
there are very significant correlations of the group size,
velocity dispersion and virial mass with $n_m$. 
A least square fit to the corrected, for the redshift-bias and
discreteness effects, data ($n_m\ge4$) gives:
$$\sigma_v/{\rm km \;s^{-1}}  \simeq 4.83 (\pm 0.26) n_m + 122 (\pm 3)$$
$$ R_v/(h^{-1}_{72} {\rm Mpc})\simeq 0.0212 (\pm 0.0008) n_m + 0.254 (\pm 0.007)$$
$$\log_{10}(M_v/M_{\odot})=0.0416 (\pm 0.0018) n_m + 12.36 (\pm 0.02) \;.$$
The above trends of the group projected size, velocity dispersion and
virial mass could be a natural consequence of the hierarchy of
cosmic structures.

It is also evident from Fig. 7 that the 2PIGG groups have
consistent crossing times, independent of the number of galaxy members,  
of $\sim 1.5$ Gyr's 
which implies that the majority of them could be virialized systems
(except probably for those formed within the last few Gyrs). 
Had we used instead of the virial radius the de-projected $r_{\rm
max}$ value we would have found a median crossing time of $\sim 3.8$
Gyrs, still significantly smaller than the age of the universe.

\section{Conclusions}
Using Monte-Carlo and N-body simulations we have
investigated the different biases that enter in the determination of
the 2PIGG morphological and dynamical characteristics 
and we have devised statistical correction
procedures to recover their bias free values.

Within a redshift that the 2PIGG groups
have a roughly constant space number density ($z\le 0.08$),
we derived the average morphological and dynamical characteristics
(size, velocity dispersion, virial mass and crossing times), which we
find to be significantly smaller than those of other recent studies,
exactly because we have taken into account the redshift-dependant
bias. 
Within $z\le 0.08$, the median value of the  group virial radius,
virial mass and crossing time is $\sim 6 \times
10^{12} \; h^{-1}_{72} M_{\odot}$, $\sim 0.4 \; h^{-1}_{72}$ Mpc and $\sim 1.5$ Gyr's, respectively.

Assuming that groups constitute a homogeneous 
spheroidal population, we numerically invert the 
projected, discreteness free,
axial ratio distribution to obtain the corresponding intrinsic 
one. The only acceptable model is that of a prolate, or a triaxial
with pronounced prolatness,
distribution with a mean axial ratio of $\beta\simeq 0.46$ and a dispersion
of $\sim 0.16$.

We have also found that there is a correlation of the size and
the velocity dispersion with the number of group members, 
which is probably a natural outcome of the hierarchy of cosmic structures.

\section* {Acknowledgements}
C.Ragone-Figueroa acknowledges financial support by the LENAC network 
and thanks INAOE for its hospitality. We also thank the referee,
T. Goto, for useful suggestions.

\begin{table}
\caption[]{The median dynamical characteristics of 2PIGG groups with
$4\le n_m \le 30$, for various cuts in redshift and
for the redshift bias corrected and uncorrected cases. We also present
the uncorrected case for the very local volume ($z\le 0.03$), were the
redshift bias is inexistent.
}
\tabcolsep 10pt
\begin{tabular}{cccccccc} \\ \hline
 $z$        &$\#$ & Correct bias& ${\overline \sigma}_v /{\rm km s}^{-1}$ &
 ${\overline M}_v /h^{-1}_{72} M_{\odot}$ & ${\overline R}_{v}/ h^{-1}_{72} {\rm
 Mpc}$ & $\overline{r}_{\rm max}/ h^{-1}_{72} {\rm Mpc}$ & ${\overline \tau}/{\rm yr}$ \\ \hline
 $\le 0.08$ & 1728  & Yes& 150$\pm 34$ & $5.7\times 10^{12}$ & 0.38 & 0.73 & $1.5 \times 10^9$ \\
 $\le 0.03$ & 199  & No & 157$\pm 35$ & $6.2\times 10^{12}$ & 0.40 & 0.75 & $1.3 \times 10^9$ \\
 $\le 0.08$ & 1728  & No & 188$\pm 45$ & $1.4\times 10^{13}$ & 0.62 & 1.05 & $1.9 \times 10^9$ \\
 $\le 0.2$  & 6128  & No & 257$\pm 70$ & $4.2\times 10^{13}$ & 0.98 & 1.71 &$2.2 \times 10^9$ \\
\hline
\end{tabular}
\end{table}

\end{document}